\documentclass[twocolumn]{aastex62}

\usepackage{amsmath}
\usepackage{graphicx}
\usepackage{color}

\usepackage{amsmath}

\newcommand{\beq}{\begin{equation}}
\newcommand{\eeq}{\end{equation}}

\newcommand\T{\rule{0pt}{2.8ex}}       
\newcommand\B{\rule[-1.4ex]{0pt}{0pt}} 

\begin{document}

\title{Interferometric Closure Phase Uncertainties in the Low Signal-to-Noise Ratio Regime}

\correspondingauthor{Pierre Christian}
\email{pierre.christian.b@gmail.com}

\author{Pierre Christian and Dimitrios Psaltis}
\affiliation{Astronomy Department, University of Arizona, 933 N Cherry Ave, Tucson, AZ 85719}

\begin{abstract}
Closure phases are critical in astronomical interferometry. However, their uncertainties are difficult to compute numerically. We provide a method to efficiently compute interferometric closure phase distributions in terms of an approximate distribution that is valid in the low signal-to-noise ratio regime. This is done by first showing that the true phase distribution is well approximated by the von Mises distribution, then performing a convolution of three von Mises distributions. The resulting approximation is superior than the normal distribution for all signal-to-noise ratios and, being fully analytic, allow for fast computations in statistical algorithms.  \\
\end{abstract}

\section{Introduction}
The closure phase is one of many closure quantities that can be defined over an interferometric array \citep{1958MNRAS.118..276J}. While the phase measurement of each baseline is laden with antenna-dependent phase noise, these noises cancel out in the closure phases, leaving an observable that is affected only by the thermal noise. Recently, closure phases have been successfully used in a variety of astronomical observations, such as probing gas and dust around massive stars \citep{2017ASPC..508..163M}, resolving rings around protoplanetary disks \citep{2017ApJ...842...77S}, detecting orbital motion close to the supermassive black hole Sgr A* \citep{GRAVITY}, as well as detecting cosmic reionization \citep{2018PhRvL.120y1301T}. At millimeter wavelengths, using closure quantities is the most promising avenue for interferometric imaging  (see, e.g., \citealt{2018ApJ...857...23C}). 

Closure phases are one of the key interferometric observables used by the Event Horizon Telescope (EHT)\footnote{https://eventhorizontelescope.org/}, a global very long baseline interferometer, to obtain horizon-scale images of supermassive black holes \citep{PaperI,PaperII,PaperIII,PaperIV,PaperV,PaperVI}. In the case of Sgr~A*, the black hole in the center of the Milky Way, for which the dynamical timescale is shorter than the duration of a single imaging observation, the time evolution of closure phases also provides a direct handle on the variability of the underlying image \citep{2009ApJ...695...59D,2016A&A...588A..57F,2017ApJ...847...55R,2018ApJ...864....7M}. 

Despite the usefulness of closure phases in astronomical interferometry, the noise statistics of the closure phase have not been fully explored. In the high signal-to-noise ratio (SNR) regime, the closure phase distribution becomes a normal distribution, because in this limit the antenna-phase distribution also approaches a normal distribution \citep{TMS}. While replacing the closure phase distribution with a normal distribution is a valid strategy for interferometric observations at high SNR, the normal distribution is generally a poor representation of the full closure phase distribution. For example, the closure phase distribution is defined on a circle and thus, if seen as a function over the real line, it possesses periodicity. Being a square integrable function, it is impossible for the normal distribution to replicate this feature.

These deviations from the normal distribution become important in regions where the observation target possesses low SNR, for example close to minima in the interferometric (u-v) space. In the EHT, detecting these minima are important as they encode the size of the black hole shadow \citep{2018ApJ...864....7M, PaperI,PaperVI}. However, because of the numerical cost of integrating error functions, the uncertainties in the measurement of closure phases are incorporated in, e.g., the Markov Chain Monte Carlo algorithms employed using the approximate normal distribution \citep{PaperVI}. This approximation might, therefore, introduce biases in the statistical inferences related to the properties of black hole shadows.

In this article, we propose an approximation scheme that allows for the full closure phase distribution to be explored in the low SNR regime in full analyticity. In particular, in Section 2, we demonstrate that the von Mises distribution accurately approximates the full phase distribution of individual stations. In Section 3, we derive our approximations to the closure phase distribution that are valid throughout all SNR regimes. Finally, in Section 4, we present our conclusions.

\section{von Mises approximation to the phase distribution}
While in the high SNR regime the distribution of the argument of the complex phase of a baseline approaches a normal distribution, the full distribution over phase is given by \citep{TMS}
\begin{align} \label{eq:fullphase}
&P(\phi) = \frac{e^{-\frac{\textrm{SNR}^2}{2}}}{4 \pi  \sigma } \left( \sqrt{2 \pi } \textrm{SNR} \cos (\phi ) e^{\frac{\textrm{SNR}^2 \cos ^2(\phi )}{2}}  \right. \nonumber
\\&\hspace{80pt} \times \left. \left\{ \text{Erf}\left[\frac{\textrm{SNR} \cos (\phi )}{\sqrt{2}  }\right]+1\right\} +2 \right) \; , 
\end{align}
with the signal-to-noise ratio (SNR) given by $\textrm{SNR}=V/\sigma$. This distribution is expensive to compute numerically, a problem that is exacerbated by the fact that the closure phase distribution is a convolution of three such distributions. As such, we seek a distribution that can approximate the true phase distribution in the low SNR regime.

\begin{figure}
\centering
\includegraphics[width=3.3in]{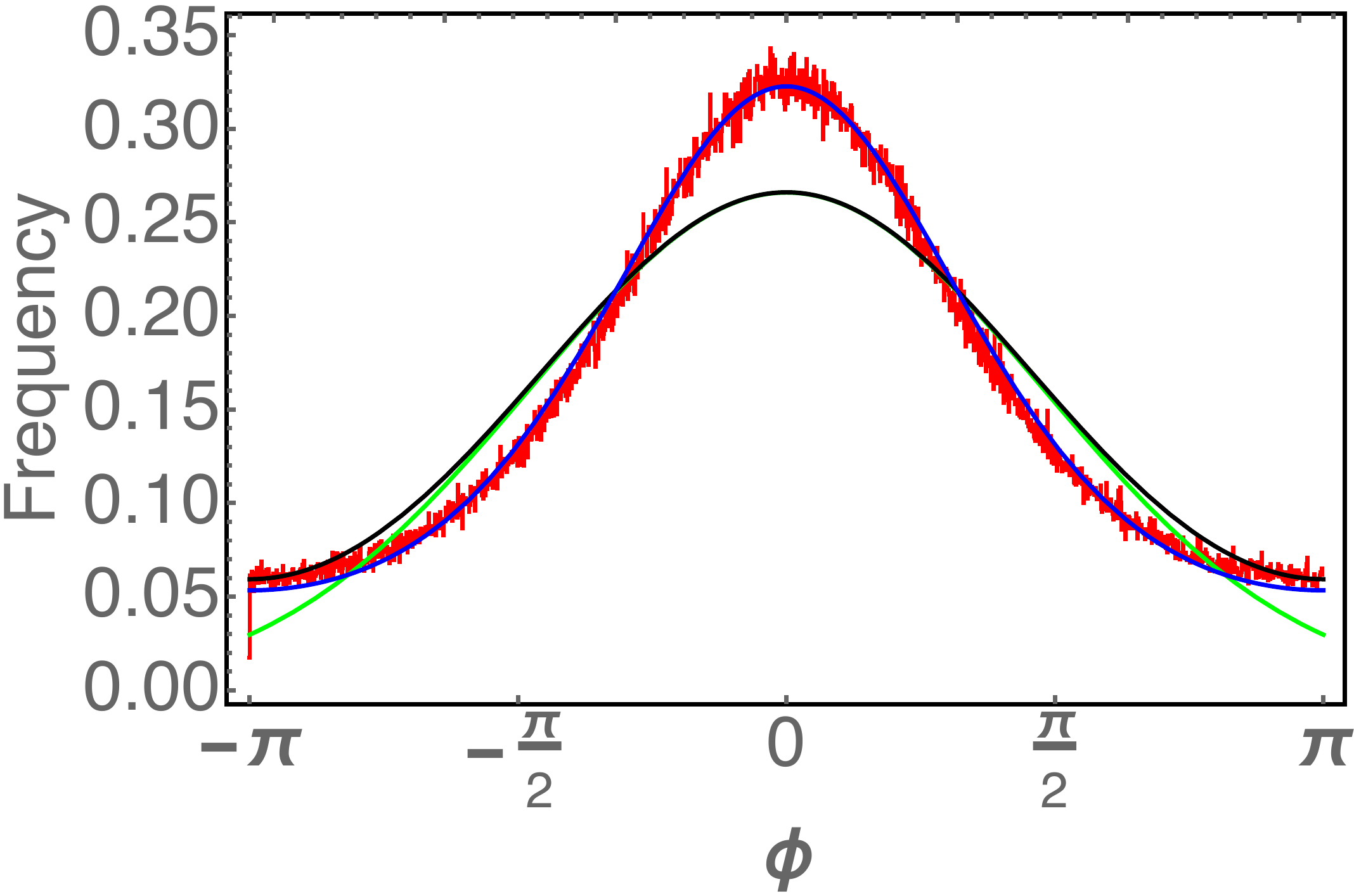}
\caption{The wrapped normal distribution (black), the normal distribution (green), and the von Mises distribution (blue) plotted along with a histogram of the true phase distribution (red) for $\sigma=1.5$. The von Mises distribution provides a superior approximation at low SNRs.}
\label{fig:wn}
\end{figure}

A first attempt to approximate the distribution described by equation (\ref{eq:fullphase}) is by the wrapped normal distribution,
\beq
\textrm{WN}(\mu, \sigma, x) = \frac{1}{\sigma \sqrt{2\pi}} \sum_{k=-\infty}^{\infty} \exp \left[ \frac{- (\theta - \mu + 2 \pi k)^2}{2 \sigma^2} \right] \; ,
\eeq
where $\mu$ is the mean of the distribution and $\sigma$ plays an analogous role to the standard deviation of a standard normal distribution, with the variance of $WN(\mu,\sigma,x)$ given by $(1 - e^{-\sigma^2/2})$. However, the wrapped normal distribution fails to capture the true phase distribution at low SNRs (c.f. Figure \ref{fig:wn}). 

The true phase distribution is better captured by the von Mises distribution,
\beq
\mathrm{vM} (\mu, \kappa, x) \equiv f_M(x | \mu, \kappa) = \frac{e^{\kappa \cos(x - \mu)}}{2 \pi I_0(\kappa)} \; ,
\eeq
where $\kappa$ is the concentration parameter and $\mu$ the location parameter (see also Figure Figure \ref{fig:wn}). Unlike the normal distribution, both the wrapped normal distribution and the von Mises distribution are defined on a circle and are more apt to describe periodic random variables \citep{Fisher,Pewsey,Ley}.

\begin{figure}
\centering
\includegraphics[width=3.3in]{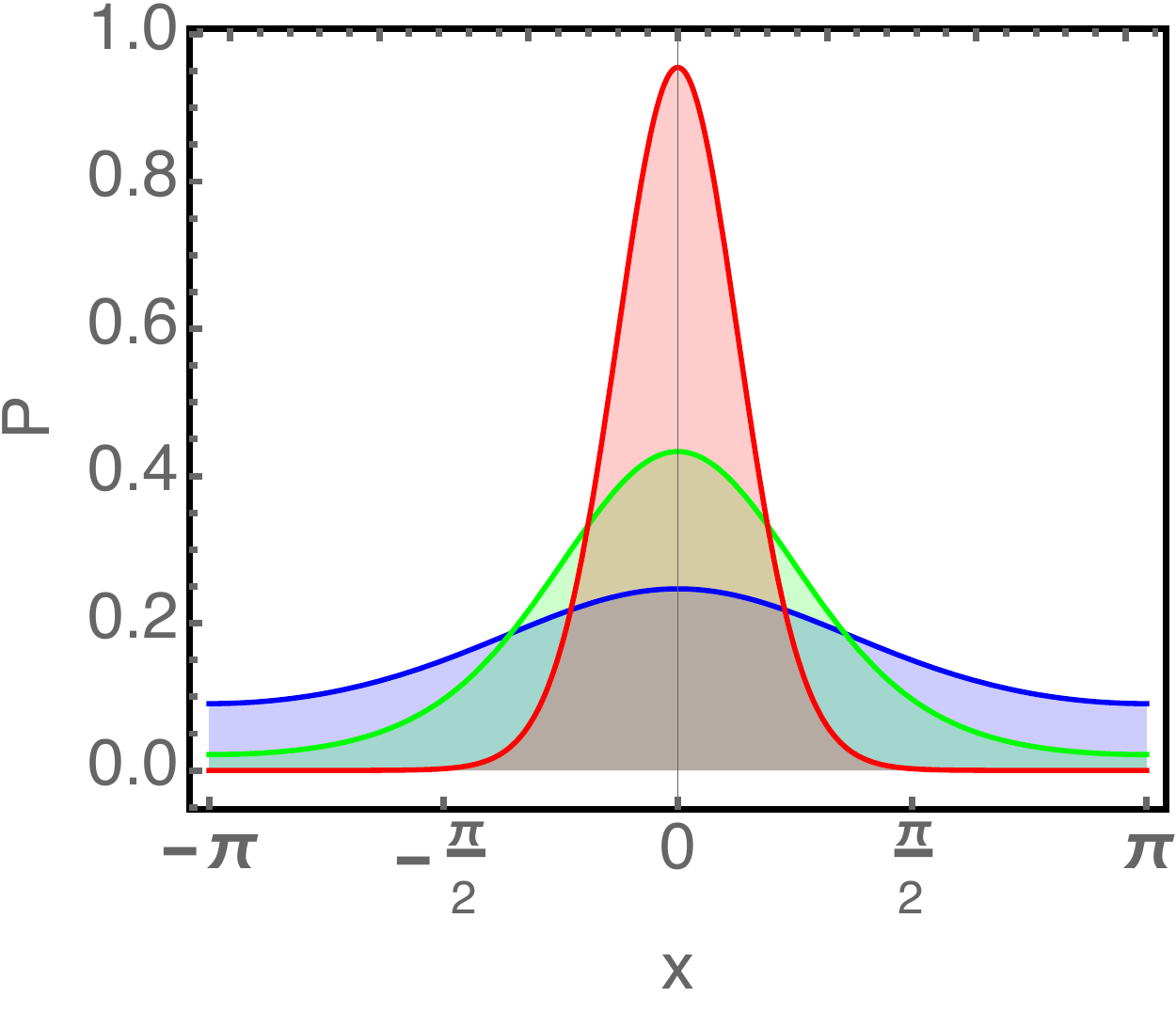}
\caption{Von Mises distributions with concentration parameter $\kappa=0.5$ (blue), $\kappa=1.5$ (green), and $\kappa=4$ (red). For all concentration parameters, we set the mean to $\mu=0$.}
\label{fig:vonmises}
\end{figure}

In analogy to the normal distribution, the concentration parameter $\kappa$ of the von Mises distribution is related to the SNR. Similarly, the location parameter $\mu$ is the mean and median of the distribution, which, without loss of generality, will be set to $0$ in this work. Figure \ref{fig:vonmises} shows the von Mises distribution for a variety of $\kappa$ values.

In order to find the von Mises distribution that corresponds to a particular value of the SNR in equation (\ref{eq:fullphase}), we first expand the true expression for the distribution over phases to second order in $\phi$ as 
\beq
P(\phi) = P_0(\phi) + P_2(\phi) \phi^2 + O[\phi^3] \; ,
\eeq
with 
\beq
P_0(\phi) = \frac{\sqrt{2 \pi } \textrm{SNR} \left[\text{Erf}\left(\frac{\textrm{SNR}}{\sqrt{2}  }\right)+1\right]+2   e^{-\frac{\textrm{SNR}^2}{2 }}}{4 \pi  }  \;,
\eeq
and
\begin{align}
P_2(\phi) =& -\frac{\textrm{SNR}}{8 \pi } \Bigg\{\sqrt{2 \pi } \left(1+\textrm{SNR}^2\right) \Big. \nonumber 
\\ &\quad\quad \Big. \times \left[\text{Erf}\left(\frac{\textrm{SNR}}{\sqrt{2}  }\right)+1\right]  +2   \textrm{SNR} e^{-\frac{V^2}{2 }} \Bigg\}  \; .
\end{align}

Taking the ratio of the second to the zeroth order term, we obtain
\begin{align}
\frac{P_2(\phi)}{P_0(\phi) } &= \frac{\textrm{SNR}}{2  \left\{ \sqrt{2 \pi } \textrm{SNR} e^{\frac{\textrm{SNR}^2}{2 }} \left[\text{Erf}\left(\frac{\textrm{SNR}}{\sqrt{2} }\right)+1\right]+2 \right\}} \nonumber 
\\&\quad\quad\quad\times \Big\{ \sqrt{2 \pi } e^{\frac{\textrm{SNR}^2}{2 }} \left(1 +\textrm{SNR}^2\right) \big. \nonumber
\\&\quad\quad\quad\times \big. \left[\text{Erfc}\left(\frac{\textrm{SNR}}{\sqrt{2}  }\right)-2\right]-2  \textrm{SNR} \Big\}  \; .
\end{align}

Doing the same to the von Mises distribution,
\beq
\mathrm{vM}(0, \kappa, \phi) = \mathrm{vM}_0(\phi) + \mathrm{vM}_2(\phi)\phi^2 + O[\phi^3] \;,
\eeq
with
\beq
\mathrm{vM}_0 (\phi) = \frac{e^{\kappa}}{2 \pi I_0(\kappa)} \;,
\eeq
and
\beq
\mathrm{vM}_2(\phi) = - \frac{\kappa \phi^2 e^\kappa}{4 \pi I_0(\kappa)} \;,
\eeq
we obtain
\beq
\frac{\mathrm{vM}_2(\phi)}{\mathrm{vM}_0 (\phi)} = -\frac{\kappa}{2} \; .
\eeq
Setting 
\beq
\frac{P_2(\phi)}{P_0(\phi} = \frac{\mathrm{vM}_2(\phi)}{\mathrm{vM}_0(\phi)} \; ,
\eeq
 we obtain the relationship between the von Misses concentration parameter $\kappa$ and $\textrm{SNR} = V/\sigma$, 
\begin{align} \label{eq:kappa}
\kappa(\textrm{SNR}) &= - \frac{\textrm{SNR}}{\left\{ \sqrt{2 \pi } \textrm{SNR} e^{\frac{\textrm{SNR}^2}{2 }} \left[\text{Erf}\left(\frac{\textrm{SNR}}{\sqrt{2}  }\right)+1\right]+2  \right\} } \nonumber
\\&\quad\quad \times \Bigg\{ -2 \textrm{SNR} + \sqrt{2 \pi } e^{\frac{\textrm{SNR}^2}{2 }} \left(1+\textrm{SNR}^2\right) \nonumber 
\\&\quad\quad \times \left[\text{Erfc}\left(\frac{\textrm{SNR}}{\sqrt{2} }\right)-2\right] \Bigg\} \; .
\end{align}

By asymptotically matching the limits of equation (\ref{eq:kappa}), we obtain a first approximation for $\kappa(\textrm{SNR})$, which could be used to facilitate faster computations,
\beq \label{eq:kappaapprox}
\kappa(\textrm{SNR}) \approx \frac{\sqrt{\frac{\pi} {2}} \textrm{SNR}}{1+ \sqrt{\frac{\pi} {2}} \textrm{SNR}} + \textrm{SNR}^2  \;. 
\eeq

The accuracy of using $\textrm{vM}(\mu, \kappa(\textrm{SNR}),\phi)$ with $\kappa(\textrm{SNR})$ given by equation (\ref{eq:kappa}) to approximate the true phase distribution is worst at around SNR$\sim 1$ (see Appendix A for a quantitative measure of the accuracy of the von Mises approximation). Therefore, we can improve this approximation by performing an ad hoc correction that makes the approximate distribution behave more like the true phase distribution in this regime. One way to perform this correction is to add a corrective term to $\kappa$ at around an SNR of unity,
\begin{align} 
\kappa'(\textrm{SNR}) &= \kappa(\textrm{SNR}) - H\left( \frac{1}{\textrm{SNR}}-0.5 \right) \\
&\approx \frac{\sqrt{\frac{\pi} {2}} \textrm{SNR}}{1+ \sqrt{\frac{\pi} {2}} \textrm{SNR}} + \textrm{SNR}^2   - H\left( \frac{1}{\textrm{SNR}}-0.5 \right) \; , \label{eq:kappacorrect}
\end{align}
where $H$ is the Hamming window, 

 \[ H(x) \equiv \begin{cases} 
      \frac{25}{46} + \frac{21}{46} \cos \left( 2 \pi x  \right) &-\frac{1}{2} \le x \le \frac{1}{2} \\
      0 & |x| > \frac{1}{2} \; . 
   \end{cases}
\]
This choice of windowing function is not unique, and different windows can be used equivalently. By performing the correction on $\kappa$ instead of modifying the shape of the distribution itself, our approximate distribution remains the von Mises distribution, but one whose concentration parameter is mapped from a different SNR than the $\kappa(\textrm{SNR})$ given by moment matching as in equation (\ref{eq:kappa}).

\begin{figure}
\centering
\includegraphics[width=3.2in]{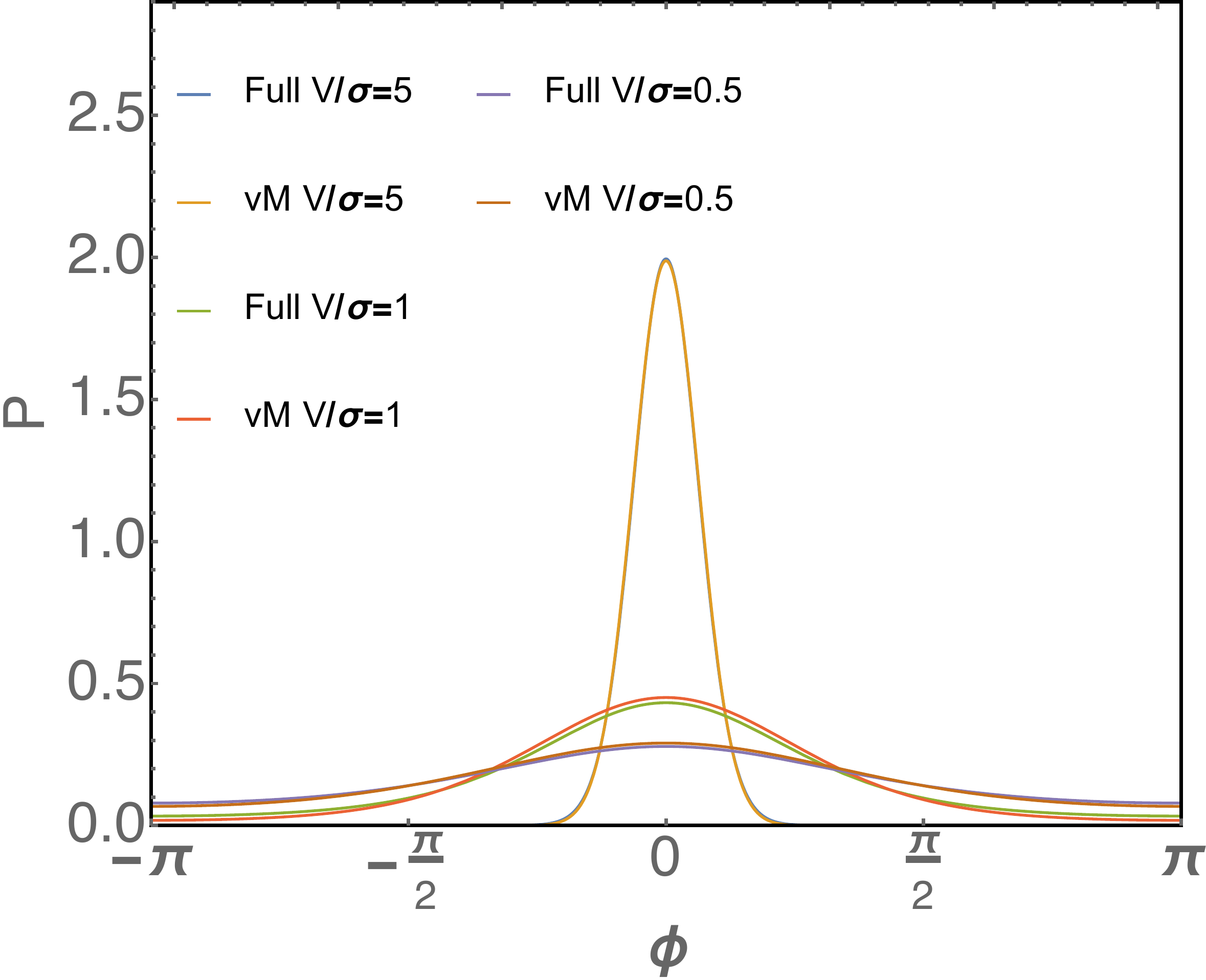}
\caption{Comparison between the full phase distribution, equation (\ref{eq:fullphase}), with the appropriate von Mises approximations for a variety of $\textrm{SNR}=V/\sigma$. For all SNRs, we set $\mu=0$.}
\label{fig:vMvsFull}
\end{figure}

For further computational practicality, we cite the following approximation for the error function, which has an error bounded to a maximum of $1.5 \times 10^{-7}$ \citep{AandSbook},
\beq \label{eq:erfapprox}
\rm{Erf}(x) \approx 1 - (a_1 t + a_2 t^2 + \ldots + a_5 t^5) e^{-x^2} \; ,
\eeq
where 
\beq
t = \frac{1}{1 + px} \; ,
\eeq
with the numerical constants,
\begin{align}
p&=0.3275911 \; , \nonumber \\ a_1&=0.254829592 \; , \nonumber \\ a_2&=-0.284496736 \; , \nonumber \\ a_3 &= 1.421413741 \; , \nonumber \\ a_4 &= -1.453152027  \; , \nonumber\\ a_5&=1.061405429 \; .\nonumber
\end{align}
In Figure (\ref{fig:vMvsFull}) we plot the full phase distribution, equation (\ref{eq:fullphase}) with the appropriate von Mises approximations for a variety of $ \textrm{SNR}=V/\sigma$. The error is $\sim 10$\% for low SNR and improves as SNR is increased. By $\mathrm{SNR}\sim3$, the error is at the few percent level. A comparison between the von Mises approximation to a previously known approximation of the phase distribution in the noise-dominated regime \citep{1976MExP...12..228M,TMS} is provided in Appendix A.

\section{Approximations of the closure phase distribution}
In an interferometric array, a baseline $i$ measures the phase 
\beq
\Phi_i = \phi_i + \eta_i \; ,
\eeq
where $\phi_i$ is the phase due to the source and $\eta_i$ is the noise contribution to the phase. The closure phase is then given by,
\begin{align}
g &= \Phi_1 + \Phi_2 - \Phi_3 \nonumber 
\\&= \phi_1 + \phi_2 - \phi_3 \; ,
\end{align}
where each $\phi_i$ is a number randomly distributed according to equation (\ref{eq:fullphase}). The true closure phase distribution is then given by the convolution,
\begin{align} \label{eq:Cconvolved}
&C(x) \equiv \int_0^{2\pi} \int_0^{2\pi} P(V_1, \sigma_1, \eta_1 , \theta) \nonumber \\
&\hspace{20pt} \times P(V_2, \sigma_2, \eta_2 ,\phi-\theta) P(V_3, \sigma_3, \eta_3,x-\phi) \mathrm{d}\theta \mathrm{d} \phi \; ,
\end{align}
where $P$ is the true phase distribution given in equation (\ref{eq:fullphase}), each of which depends on the $\textrm{SNR}=V/\sigma$ and the peak phase, $\eta$, which is set to $0$ in equation (\ref{eq:fullphase}). 

Using the fact that the von Mises distribution approximates the true phase distribution, we approximate each phase distribution in the convolution with a von Mises distribution, 
\begin{align} \label{eq:Gconvolved}
&C(x) \approx G(\mu_1, \mu_2, \mu_3, \kappa_1, \kappa_2, \kappa_3, x) \equiv \int_0^{2\pi} \int_0^{2\pi} \mathrm{vM}(\mu_1, \kappa_1,\theta) \nonumber \\
&\hspace{50pt} \times \mathrm{vM}(\mu_2, \kappa_2,\phi-\theta) \mathrm{vM}(\mu_3, \kappa_3,x-\phi) \mathrm{d}\theta \mathrm{d} \phi \; ,
\end{align}
and perform the convolution in Fourier space. The Fourier series of a single von Mises distribution is given by \citep{AandSbook},
\beq
F[\mathrm{vM}]_n (\mu, \kappa) = \frac{I_n(\kappa)}{2 \pi I_0(\kappa)} \;\;\; ; \; \; \; n \in  \mathbb{Z} \; ,
\eeq
where $I_n(\kappa)$ are modified Bessel functions of the first kind. In Appendix B, we provide an approximation scheme to approximate $I_n$ in order to speed up their computations. 

In Fourier space, the convolution of three von Mises distribution is therefore 
\beq
\tilde{G}_n (\mu_1, \mu_2, \mu_3, \kappa_1, \kappa_2, \kappa_3)  = \frac{I_n(\kappa_1)I_n(\kappa_1)I_n(\kappa_3)}{(2\pi)^3 I_0(\kappa_1)I_0(\kappa_2)I_0(\kappa_3)} \; .
\eeq
In real space, the full closure phase distribution becomes
\begin{align} \label{eq:realspace}
&G(\mu_1, \mu_2, \mu_3, \kappa_1, \kappa_2, \kappa_3, x) = \nonumber \\
&\;\;\;\;\; \frac{1}{(2\pi)^3} + \sum_{n=1}^\infty \frac{I_n(\kappa_1)I_n(\kappa_1)I_n(\kappa_3)2 \cos [n(x-c_0)] }{(2\pi)^3 I_0(\kappa_1)I_0(\kappa_2)I_0(\kappa_3)} \; ,
\end{align}
where $c_0 = \mu_1+\mu_2+\mu_3$, and truncating the sum results in an approximation that is valid for low $\kappa$. We will call $G(\mu_1, \mu_2, \mu_3, \kappa_1, \kappa_2, \kappa_3, x)$ the \emph{mixture} distribution.

\begin{figure*}
\centering
\includegraphics[width=3.2in]{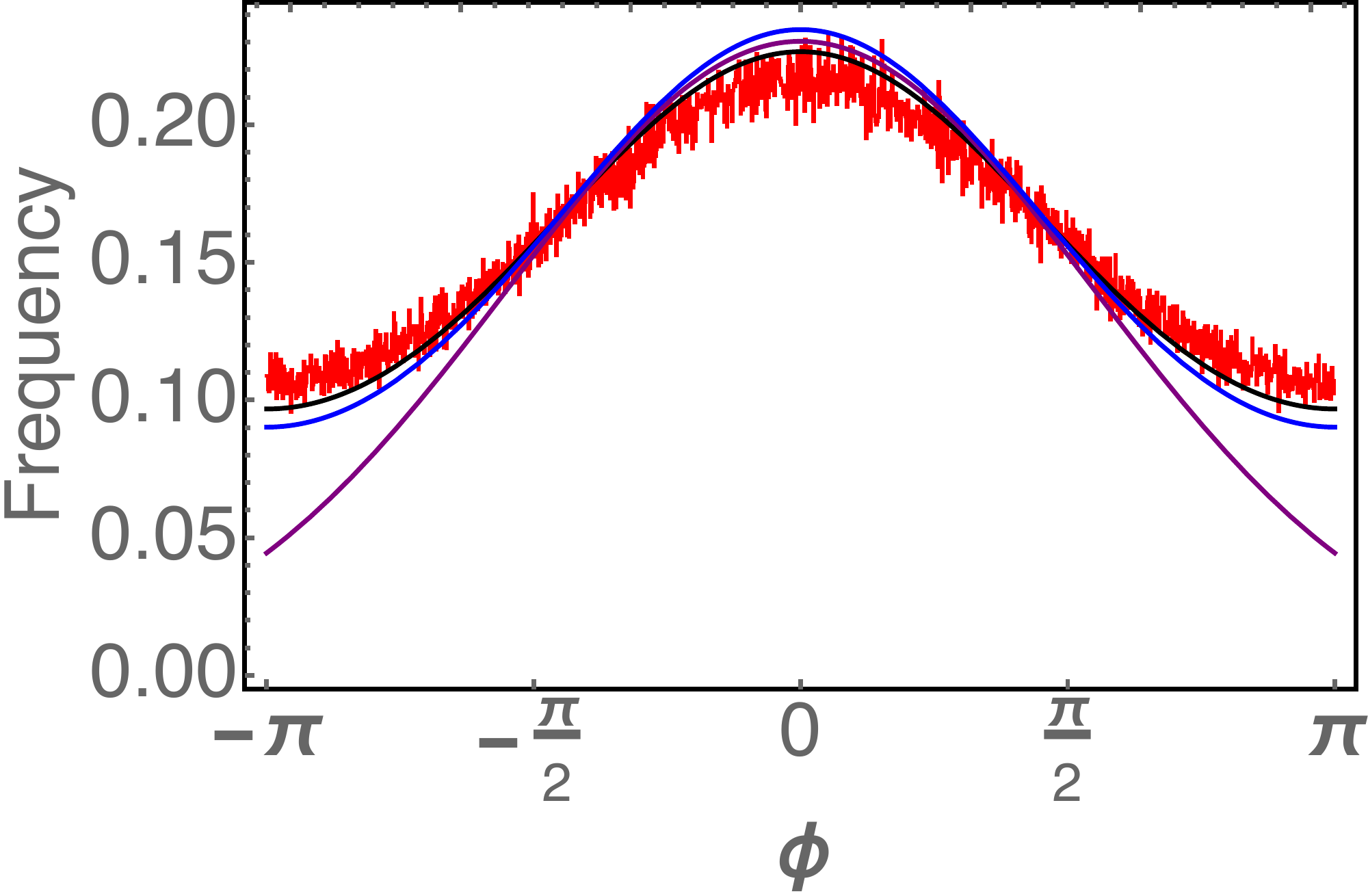}
\includegraphics[width=3.2in]{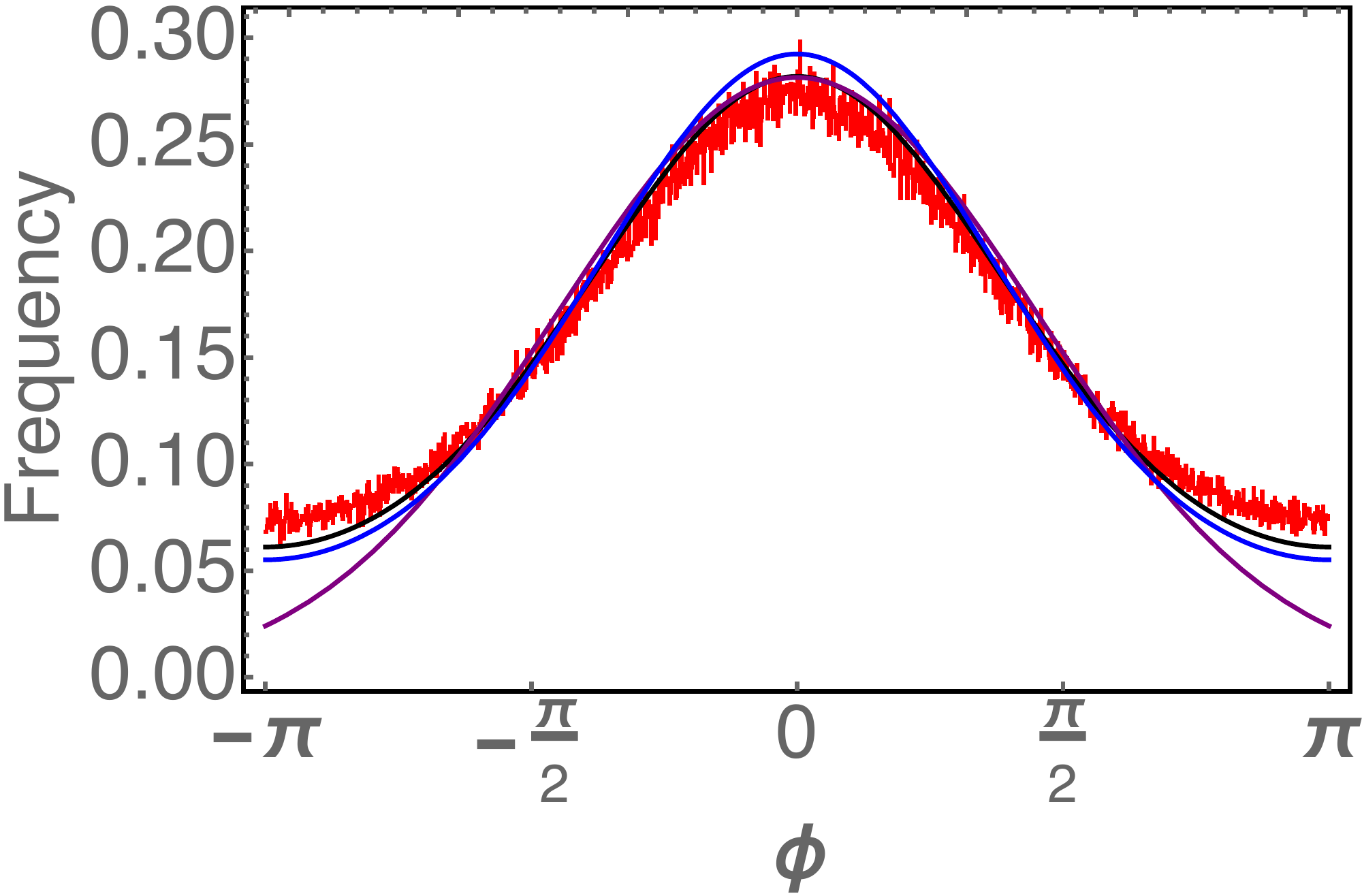}\\
\vspace{5mm}
\includegraphics[width=3.2in]{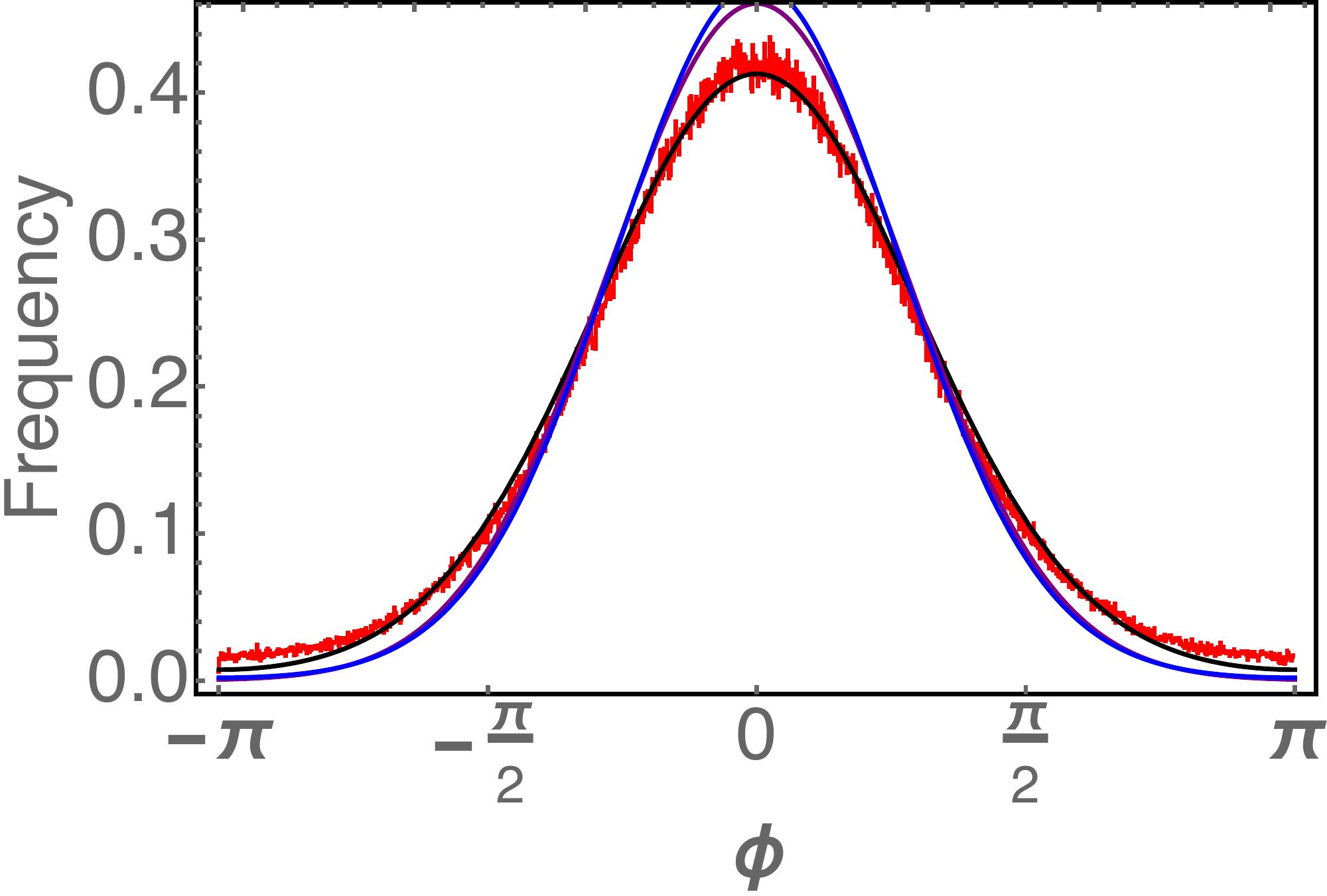}
\includegraphics[width=3.2in]{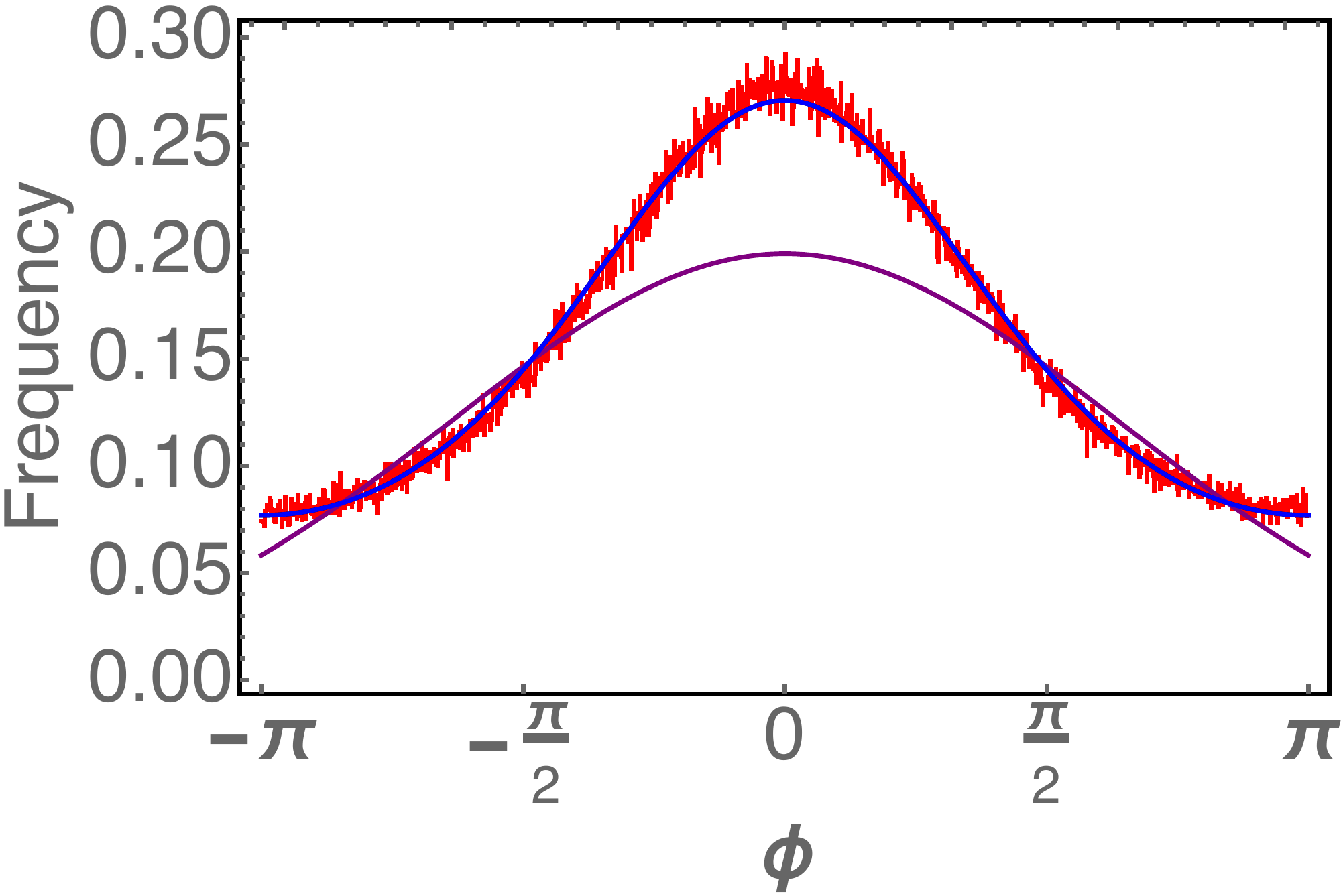}
\caption{The mixture distribution with $\kappa$ given by equation (\ref{eq:kappa}) (blue), the mixture distribution with $\kappa$ given by equation (\ref{eq:kappacorrect}) (black), the normal distribution (purple), and numerical sampling of the true closure phase distribution (red) for $(\sigma_1,\sigma_2,\sigma_3)$ set to $(1,1,1)$ (top left), $(1,1,0.1)$ (top right), $(0.5,0.5,0.5)$ (bottom left), and $(0.1,0.1,2)$ (bottom right) and $(\mu_1,\mu_2,\mu_3)$ set to $(\pi,0,0)$. The mixture distributions are truncated at the tenth order. Using the mixture distributions results in significant increase in accuracy compared to using normal distributions. In the bottom left plot, the blue and black lines are overlaid over each other. \textbf{In these plots, we have set $V=1$ so that $\textrm{SNR}=1/\sigma$.}}
\label{fig:convolve}
\end{figure*}

Figure \ref{fig:convolve} plots the mixture distribution truncated at tenth order, the normal distribution, and a histogram taken from the true closure phase distribution. As expected, at low SNRs the normal distribution fails to reproduce the true closure phase distribution. However, in this regime a tenth order truncation of equation (\ref{eq:realspace}) is an excellent approximation for the closure phase distribution.   

\begin{figure*}
\centering
\includegraphics[width=2.3in]{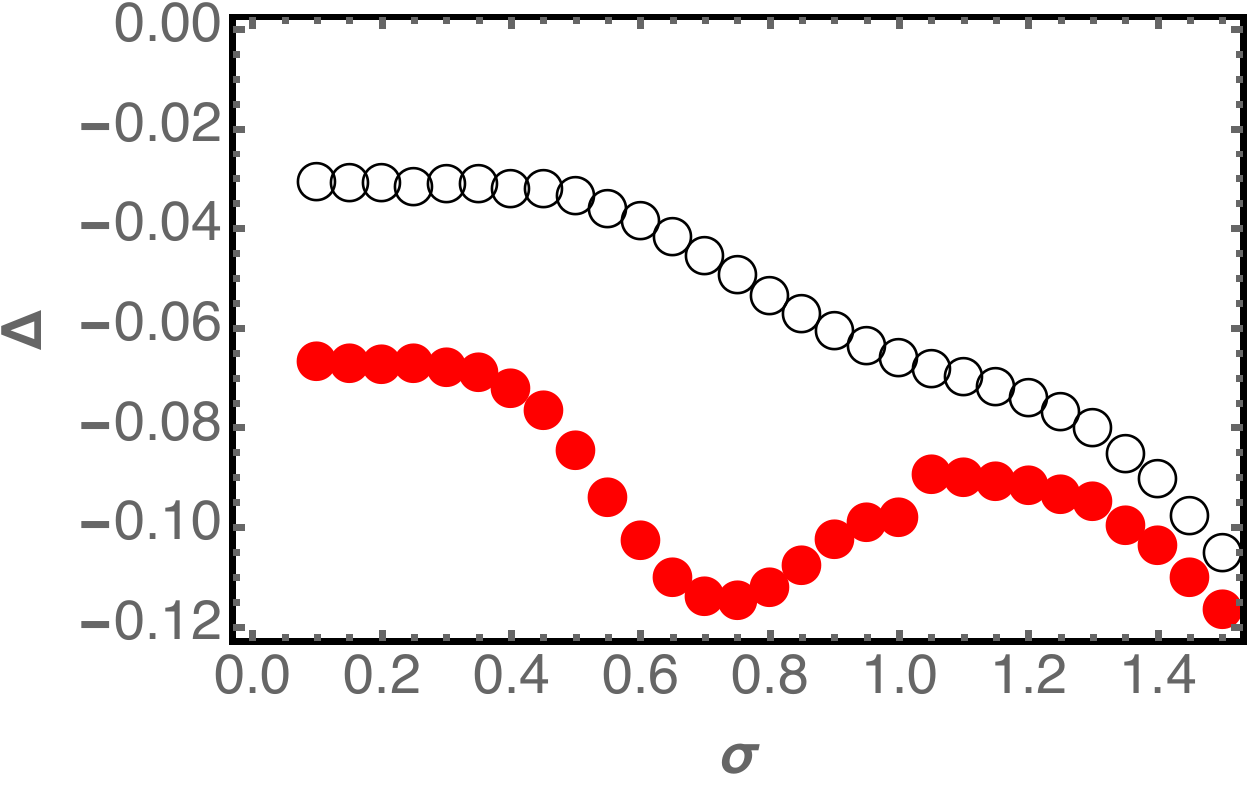}
\includegraphics[width=2.3in]{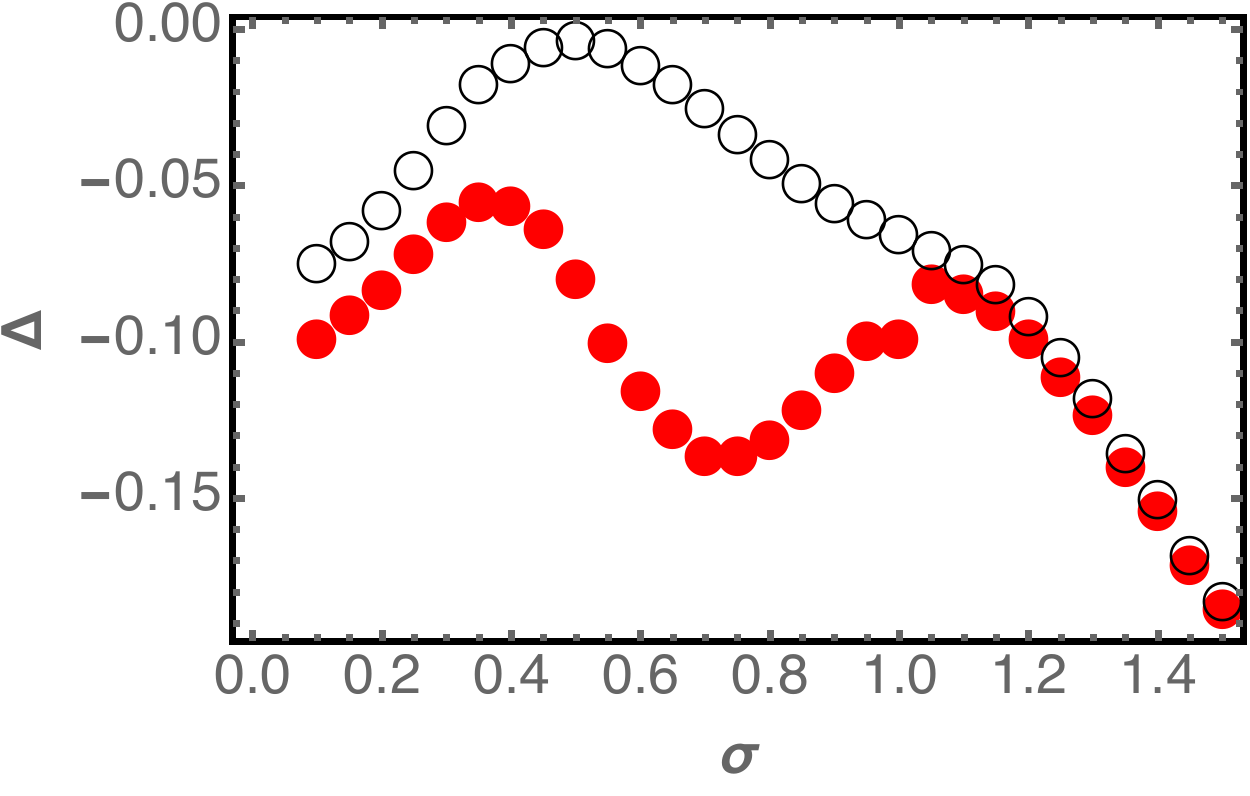}
\includegraphics[width=2.3in]{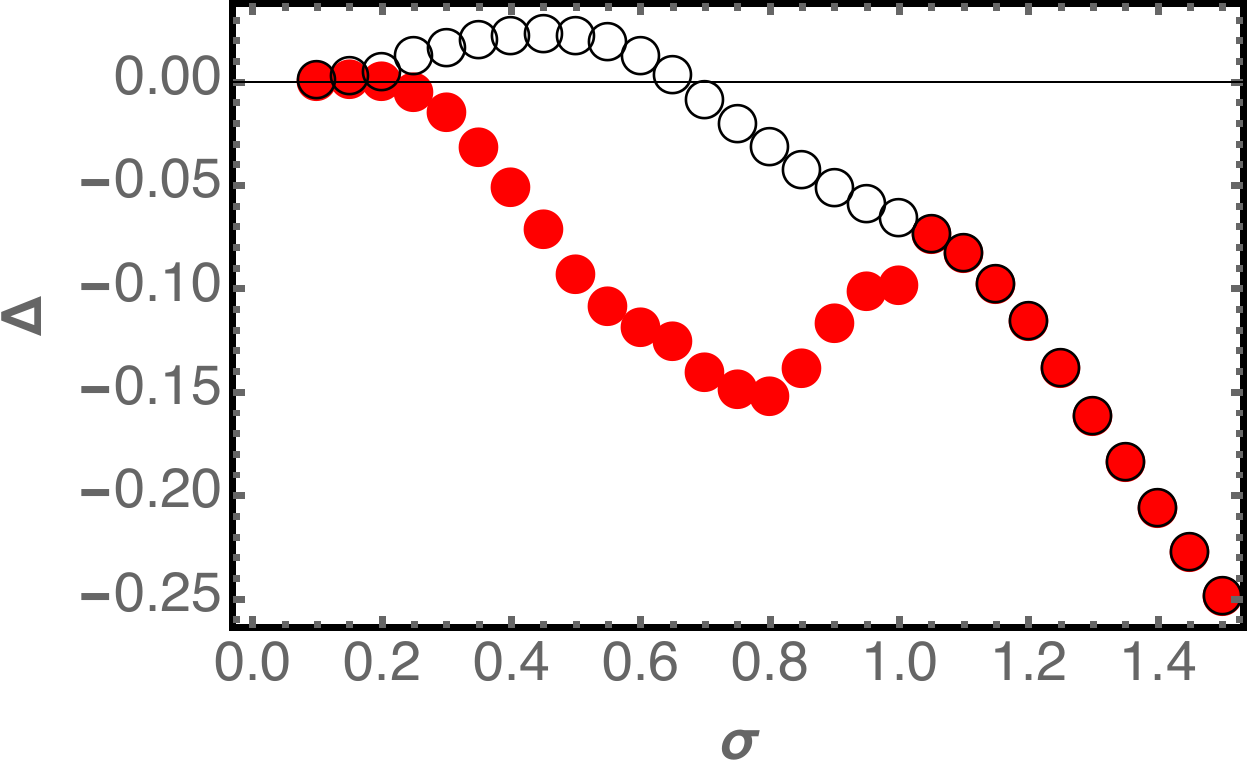}
\caption{The difference of the integrated fractional error, $\Delta$, for $(\sigma_1,\sigma_2,\sigma_3)=(\sigma,1,1)$ (\emph{left}), $(\sigma_1,\sigma_2,\sigma_3)=(\sigma,\sigma,1)$ (\emph{middle}), and $(\sigma_1,\sigma_2,\sigma_3)=(\sigma,\sigma,\sigma)$ (\emph{right}) for the mixture distribution (empty black) and the mixture distribution with the $\kappa$ correction of equation (\ref{eq:kappacorrect}) (solid red). A value of $\Delta>0$ signifies that the normal distribution more accurately represents the true closure phase distribution than the approximate distributions. \textbf{In these plots, we have set $V=1$ so that $\textrm{SNR}=1/\sigma$.}}
\label{fig:Deltas}
\end{figure*}

\subsection{Performance comparisons}
In this subsection, we will compare the performance of the mixture distribution with that of the normal distribution in approximating the full closure phase distribution. We will use two measures to evaluate their performances: the \emph{integrated fractional error} that measures how the value of the approximate distributions at each point is different from the true distribution; and the second (circular) moment, that measures how well the approximate distributions capture the wings of the true distribution.

\subsubsection{Integrated fractional error}
We define a measure for the the accuracy of an approximate distribution $Y$ by its integrated fractional error,
\beq \label{eq:intfrac}
\delta_Y = \frac{\int \left| Y(\phi) - C(\phi)\right| d\phi }{\mathrm{E}[C(\phi)]} \; ,
\eeq
where $C$ is the true closure phase distribution, $\mathrm{E}[C(\phi)]$ is the mean of $C(\phi)$, and the integral is carried out over $[0,2\pi]$. We numerically compute this value for the  mixture distribution truncated at tenth order, $\delta_{G10}$, and compare the analogous quantity for the normal distribution, $\delta_{N}$, by defining the difference of the integrated fractional error,
\beq
\Delta = |\delta_{\textrm{G10}}| - |\delta_{N}| \; .
\eeq
We plot $\Delta$ for a variety of concentration parameters in Figure \ref{fig:Deltas}. When $\Delta > 0$, the normal distribution is a better approximation to $C$ than the mixture distribution truncated to tenth order. In cases where $(\sigma_1, \sigma_2, \sigma_3) = (\sigma,\sigma,\sigma)$, the zeroth crossing is at $\sigma \approx 0.7$, while when at least one of the $\sigma$ values is of order unity, the truncation of the mixture distribution (\ref{eq:realspace}) always perform better than the normal distribution. If we use the $\kappa$ correction of equation (\ref{eq:kappacorrect}), then the truncated mixture distribution is always superior to the normal distribution except at very high SNRs, in which it becomes identical to the normal distribution, as expected. 

\subsubsection{Second moment}

\begin{figure*}
\centering
\includegraphics[width=3.2in]{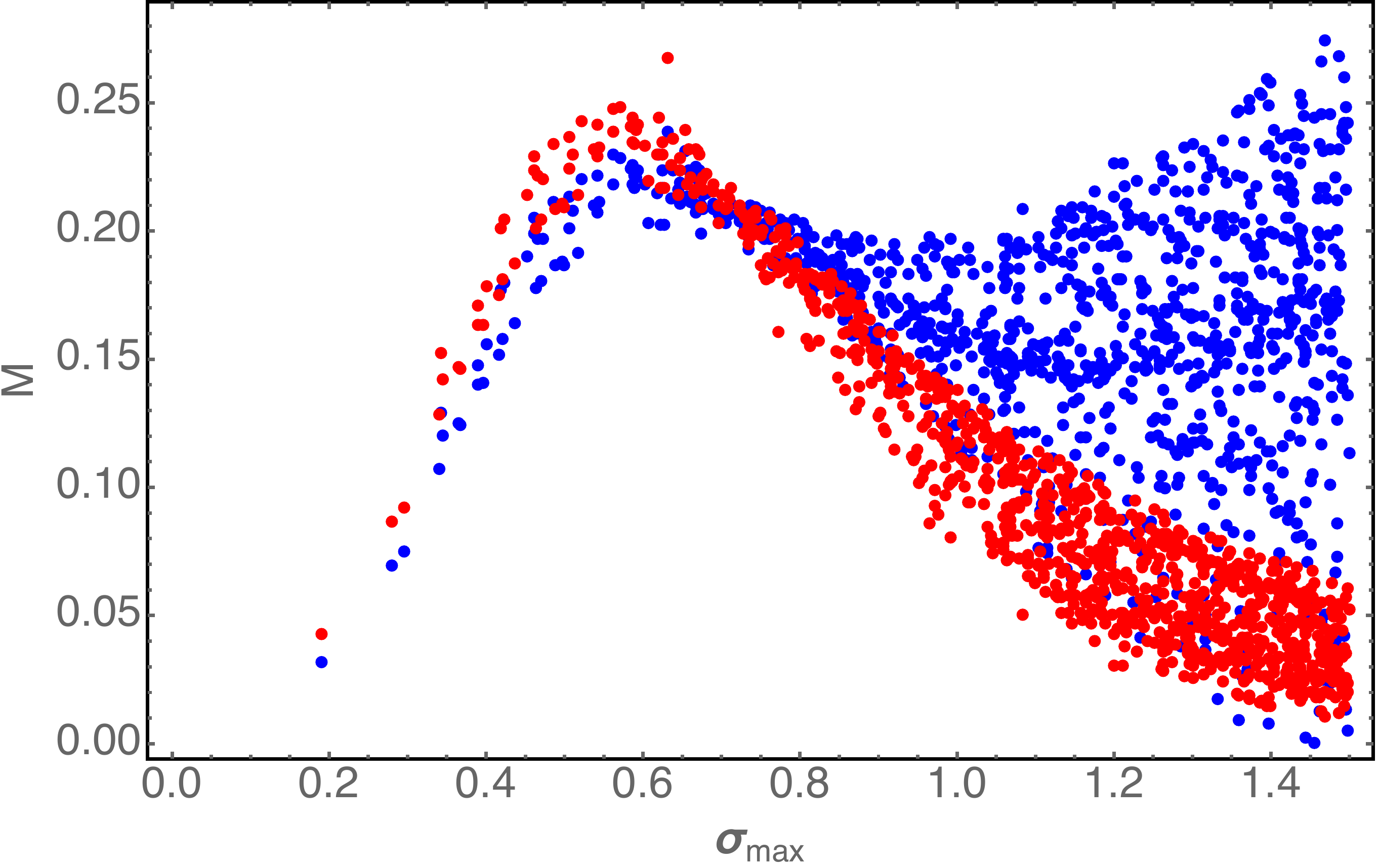}
\includegraphics[width=3.2in]{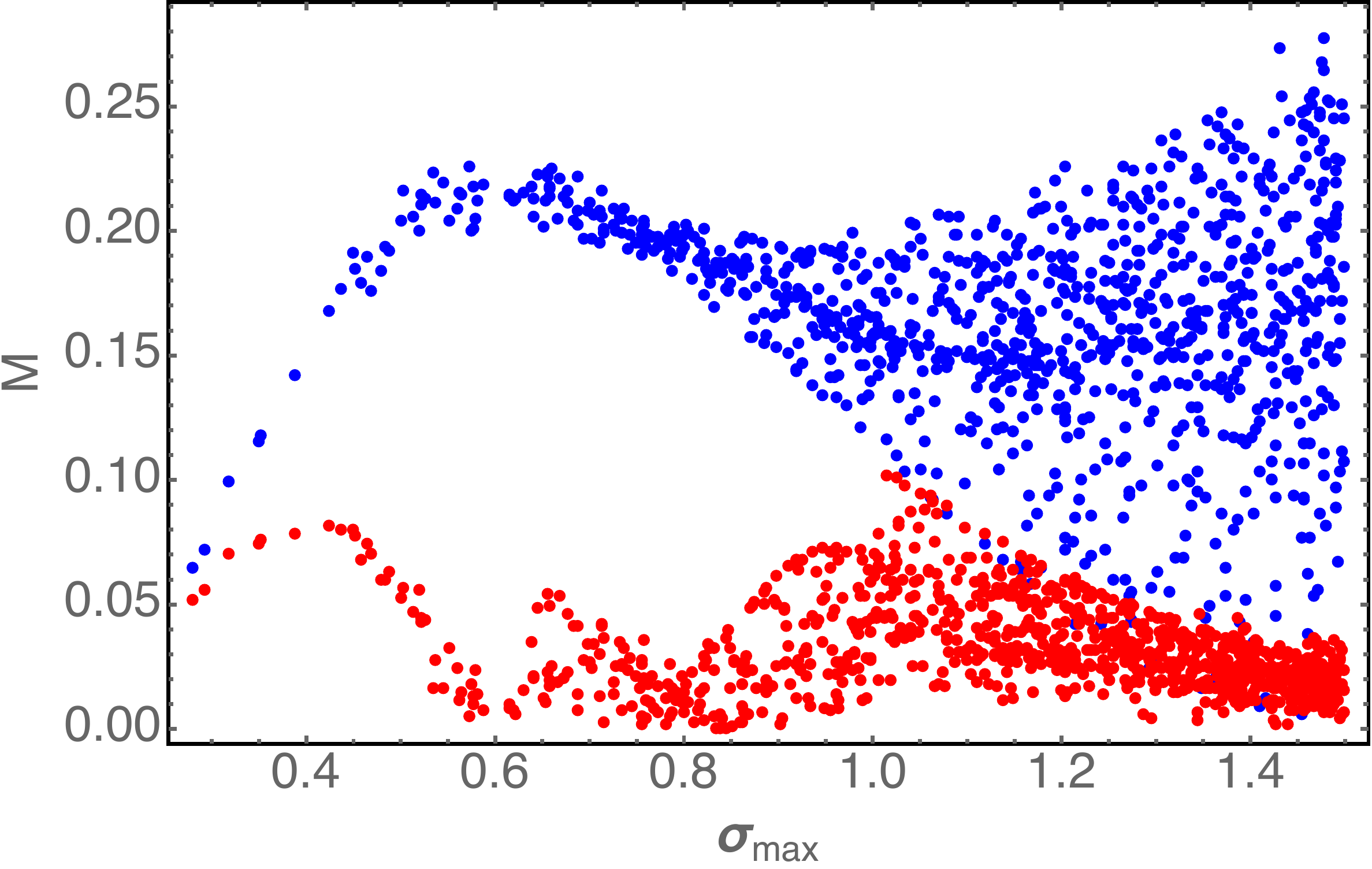}
\caption{The difference in the second circular moments of the mixture distribution truncated to tenth order (red) and the normal distribution (blue) as a function of $\sigma_{\textrm{max}}$, the maximum of $(\sigma_1,\sigma_2,\sigma_3)$. The $\sigma$'s in the plot were obtained by randomly sampling in the range $[0.1,1.5]$. The left figure shows the mixture distribution with $\kappa$ given by equation (\ref{eq:kappa}) while the right figure shows the mixture distribution with $\kappa$ given by equation (\ref{eq:kappacorrect})}
\label{fig:2ndmoment}
\end{figure*}

The normal distribution performs poorly in the low SNR regime because the true closure phase distribution develops significant non-Gaussian tails. In particular, the second moment of the normal distribution quickly diverges from that of the true closure phase distribution as the SNR is reduced. For stochastic variables defined on a circle, the second circular moment of a distribution $Y$ is given by \citep{MardiaJupp}
\beq
\mathrm{II}[Y] = \int_{-\pi}^{\pi} [1 - \cos(x-\bar{\phi})] Y \mathrm{d}x \; ,
\eeq
where $\bar{\phi}$ is the mean of the distribution. We define a measure for the performance of a distribution $Y$ in approximating the true phase distribution, $C$, as 
\beq
M[Y] = \frac{\mathrm{II}[Y] - \mathrm{II}[C]}{\mathrm{II}{C}} \; .
\eeq
In Figure \ref{fig:2ndmoment}, we plot $M[Y]$ for the convolved distribution and the normal distribution. The mixture distribution is superior to the normal distribution when $\sigma \approx 0.7$ if we used the $\kappa$ given by equation (\ref{eq:kappa}). If equation (\ref{eq:kappacorrect}) is used for $\kappa$, then the mixture distribution is superior to the normal distribution for all $\sigma$ values except when $\sigma$ is so low that the mixture distribution and the normal distribution becomes identical. 

\section{Conclusion}
As the closure phase distribution deviates from the normal distribution at low SNR, we seek for a distribution that can represent the closure phase uncertainties in this regime. Further, as this distribution will be incorporated in statistical algorithms, it is important that it is optimized and fully analytic.  

We demonstrated that the truncated mixture distribution obtained by truncating equation (\ref{eq:realspace}) along with equation (\ref{eq:kappacorrect}) to convert the SNRs of the baselines to the concentration parameter of the von Mises distribution gives an excellent approximation to the closure phase distribution  that is superior to the normal distribution for all SNRs.

In Appendix B, we provided approximations that can facilitate numerical computations of the modified Bessel functions of the first kind in the von Mises distributions. These approximations are valid at low SNR, but they remain accurate even for combination of relatively high concentration parameters, e.g., $(\kappa_1,\kappa_2,\kappa_3)=(100,100,10)$. 

In summary, the step-by-step recipe for computing the closure phase distribution using our method (up to tenth order) starting from the SNRs of the three baselines that constitute the closure phase triangle is as follows:
\begin{enumerate}
\item Use equation (\ref{eq:kappacorrect}) to convert the measured SNR for each baselines into concentration parameters.
\item Numerically compute twelve modified Bessel functions of the first kind, $I_0(\kappa_i)$, $I_1(\kappa_i)$, $I_2(\kappa_i)$, and $I_3(\kappa_i)$ with $i$ labelling individual baselines.
\item Using Table 1 in Appendix B, compute the higher order $I_n$'s.
\item Using equation (\ref{eq:realspace}) truncated to tenth order, obtain the distribution for the closure phase
\end{enumerate}

\section{Acknowledgement}
We thank the Arizona PIRE group members for useful discussions, as well as Feryal Ozel, Junhan Kim, Alan Rogers, and Jim Moran for carefully reading the manuscript. We thank the anonymous referee and the AJ statistical editor for their comments and suggestions. We gratefully acknowledge support by NSF PIRE grant 1743747.

\bibliography{BibFile.bib}

\section*{Appendix A: Comparison between the von Mises distribution and a cosine approximation in the noise dominated regime} \label{Appendix}

\begin{figure}
\centering
\includegraphics[width=3.3in]{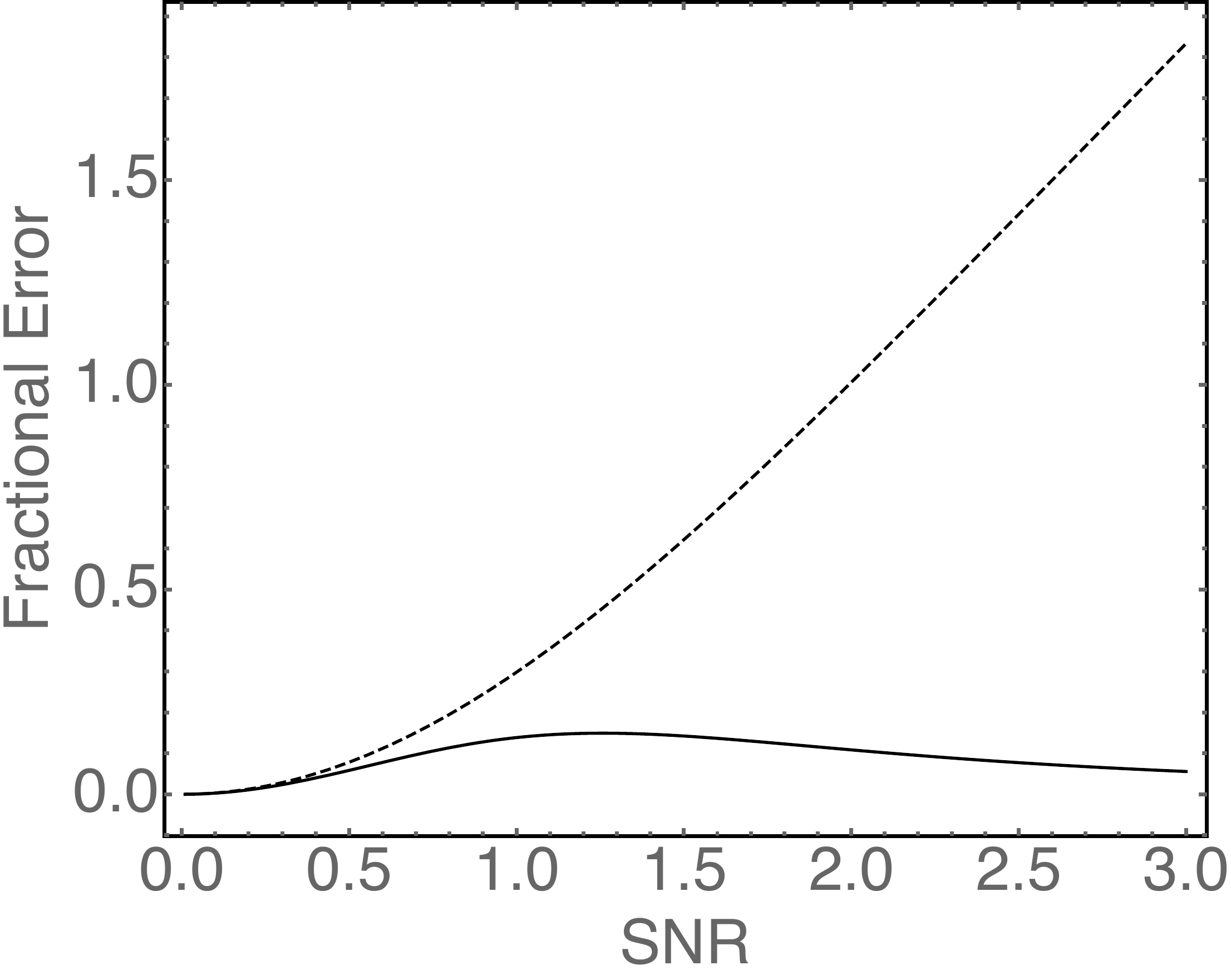}
\caption{Fractional errors of approximations of the phase distribution defined by equation (\ref{eq:approxM}) (dashed) and the von Mises distribution (solid). The von Mises distribution is superior even in the noise-dominated regime where equation (\ref{eq:approxM}) is valid.}
\label{fig:appendix}
\end{figure}

The full phase distribution has been approximated in the noise dominated regime by a constant plus a cosine function \citep{1976MExP...12..228M,TMS}, 
\beq \label{eq:approxM}
P_a(\phi) = \frac{1}{2 \pi} + \frac{\textrm{SNR}}{\sqrt{8 \pi}} \cos{\phi} \; .
\eeq
This approximation is valid when $0<\textrm{SNR}<\sqrt{2/ \pi}$. For larger SNRs, this approximation breaks down, and can produce negative probabilities. Here we compare the von Mises distribution with this approximation by computing the fractional error of this approximation as
\beq
\delta_a = \frac{\int \left| P_a(\phi) - P(\phi)\right| d\phi }{\mathrm{E}[P(\phi)]} \; ,
\eeq
and similarly the fractional error of the von Mises distribution as 
\beq
\delta_{\mathrm{vM}} = \frac{\int \left| \mathrm{vM}(\phi) - P(\phi)\right| d\phi }{\mathrm{E}[P(\phi)]} \; ,
\eeq
and plot the two fractional errors in Figure \ref{fig:appendix}. The von Mises distribution is a more accurate approximation to the phase distribution than equation (\ref{eq:approxM}) even in the latter's regime of validity. 

\section*{Appendix B:Approximations to the modified Bessel functions of the first kind} \label{Appendix B}

\begin{table*}[t] \label{Table:Coeff}
\begin{tabular}{|c|c|c|c|}
\hline
   & $I_0$                                                   & $ I_1$                                                           & $I_2$                              \T\B  \\ \hline
$I_4$ & $1+\frac{24}{k^2}   $                         & $-\frac{8}{k} - \frac{48}{k^3}           $                    &                                \T\B   \\
$I_5$ &                                                      & $1+\frac{48}{k^2}   $                                 & $-\frac{12}{k}-\frac{192}{k^3}$    \T\B \\
$I_6$ & $ 1+\frac{144}{k^2}+\frac{1920}{k^4} $ & $ -\frac{18}{k} - \frac{768}{k^3} - \frac{3840}{k^5}  $     &  \T\B \\
$I_7$ &   & $1 +  \frac{240}{k^2} + \frac{5760}{k^4} $  & $ -\frac{24}{k} - \frac{1920}{k^3} - \frac{23040}{k^5} $  \T\B \\ 
$I_8$ & $1 + \frac{480}{k^2} + \frac{28800}{k^4} + \frac{322560}{k^6} $ & $ -\frac{32}{k} - \frac{4800}{k^3} - \frac{138240}{k^5} - \frac{645120}{k^7} $ &  \T\B \\
$I_9$ &   & $1 +  \frac{720}{k^2} + \frac{67200}{k^4} + \frac{1290240}{k^6}$  & $ -\frac{40}{k} - \frac{9600}{k^3} - \frac{483840}{k^5} - \frac{5160960}{k^7}$   \T\B \\ 
$I_{10}$ & $1 + \frac{1200}{k^2} + \frac{201600}{k^4} + \frac{9031680}{k^6} + \frac{92897280}{k^8} $ & $ -\frac{50}{k} - \frac{19200}{k^3} - \frac{1693440}{k^5} - \frac{41287680}{k^7} - \frac{185794560}{k^9} $ &  \T\B \\  
 \hline 
 \end{tabular}
 \caption{Coefficients for the expansions of $I_n$ in equation (\ref{eq:Inexpansion}) up to tenth order.}
\end{table*}

In order to accelerate computations of the three convolved von Mises distributions, we provide in this section two methods to approximate the evaluations of the modified Bessel functions of the first kind. The first method is to expand the Bessel functions in terms of their polynomial expansions,
\beq
I_\nu (z) = \left( \frac{1}{2} z \right)^\nu \sum^{\infty}_{k=0} \frac{\left( \frac{1}{4} z^2 \right)^k}{k! \; \Gamma(\nu + k +1)} \; ,
\eeq
where $\Gamma(x)$ is the Gamma function. Owing to the fact that $\nu = n \in \mathbb{Z}$, we can rewrite this equation to be
\beq
I_n (z) = \left( \frac{1}{2} z \right)^n \sum^{\infty}_{k=0} \frac{\left( \frac{1}{4} z^2 \right)^k}{k! \; (n + k)!} \; .
\eeq
As this is an expansion over $z=\kappa$, truncating the expansion at some order $k$ is valid for low concentration parameters $\kappa$, i.e. when the SNR is low. Indeed, this is the regime in which we want to use the convolved von Mises distributions, as in the high SNR regime the distribution can be well approximated, in principle, by the normal distribution. 

In order to further speed up the computation of the von Mises distributions, we also present an approximation to the modified Bessel functions of the first kind that allows higher order Bessel functions to be written in terms of lower order Bessel functions. By doing so, one only needs to compute the low order Bessel functions and the higher order Bessel functions will be obtained automatically. 

First, notice that the polynomial expansion of a Bessel function of order $n$ looks like
\beq \label{eq:Inexpansion}
I_n (z) =C_{n,n} z^n + C_{2n,n} z^{2n} + C_{4n,n} z^{4n} + \ldots \; , 
\eeq
where $C_{j,n}$ denotes the coefficient of order $j$ in the expansion
\beq
C_{j,n} = \frac{1}{2^j} \left[ \left( \frac{j-n}{2} \right)! \; \left(\frac{j+n}{2} \right)!  \right]^{-1} \; .
\eeq  
By matching the first few coefficients of a higher order Bessel function with that of lower order Bessel functions, it is possible to formulate an approximation of higher order Bessel functions using lower order Bessel functions. 

As an example, for the case of $n=4$, we have
\beq
I_4(z) = \frac{z^4}{384} + \frac{z^6}{7680} + \frac{z^8}{368640} + \frac{z^{10}}{30965760} + \ldots \; .
\eeq 
We seek an approximation to $I_4(z)$, which we write as $\tilde{I}_4(z)$ with the form
\beq
\tilde{I}_4(z) = A I_0(z) + B \frac{I_1(z)}{z} + C \frac{I_0(z)}{z^2} + D \frac{I_1(z)}{z^3} \; ,
\eeq
where $(A, B, C, D)$ are coefficients that need to be solved in order to match the coefficients $C_{j,4}$. Expanding both sides, we obtain the following system of equations 
\begin{align}
\frac{A}{4}+\frac{B}{16}+\frac{C}{64}+\frac{D}{384} &= 0 \; , \nonumber 
\\ A+\frac{B}{2}+\frac{C}{4}+\frac{D}{16} & =0 \;, \nonumber
\\ C + \frac{D}{2} &= 0 \;  , \nonumber
\\ \frac{A}{64}+\frac{B}{384}+\frac{C}{2304}+\frac{D}{18432} &= \frac{1}{384} \; , \nonumber 
\end{align}
which can be solved to obtain $A=1$, $B=-8$, $C=24$, and $D=-48$. While these coefficients were matched only up to the four lowest orders of $z$, this turns out to be sufficient to approximate $I_4(z)$ up to arbitrary order, as 
\begin{align}
&A C_{j,0} + B C_{j+1,1} + C C_{j+2,0} + D C+{j+3,1} = \nonumber 
\\&\;\;\;\;\;\;\;\;\;\;\;\;\;\;\;\;\;\;\;\; \frac{1}{2^j} \left[ \left( \frac{j-4}{2} \right)! \; \left(\frac{j+4}{2} \right)!  \right]^{-1} = C_{j,4} \; .
\end{align}
In other words, solving for $(A, B, C, D)$ gives
\beq
I_4(z) \sim \tilde{I}_4(z) \; ,
\eeq 
where $\sim$ denotes that the approximation is up to arbitrary order. When the expansion is taken to infinite orders of $z$, the approximation becomes an equivalence, $I_4(z)=\tilde{I}_4(z)$. In practice, however, the order has to be truncated. This leads to an error in the very last term of $\tilde{I}_4(z)$ that becomes significant only for large $z$. This is not consequential, since, as $z=\kappa$, it is possible to simply switch to a normal approximation at large $z$, and only use the von Mises convolutions in the regime where this error is always insignificant.  

\begin{figure}[t]
\centering
\includegraphics[width=3.2in]{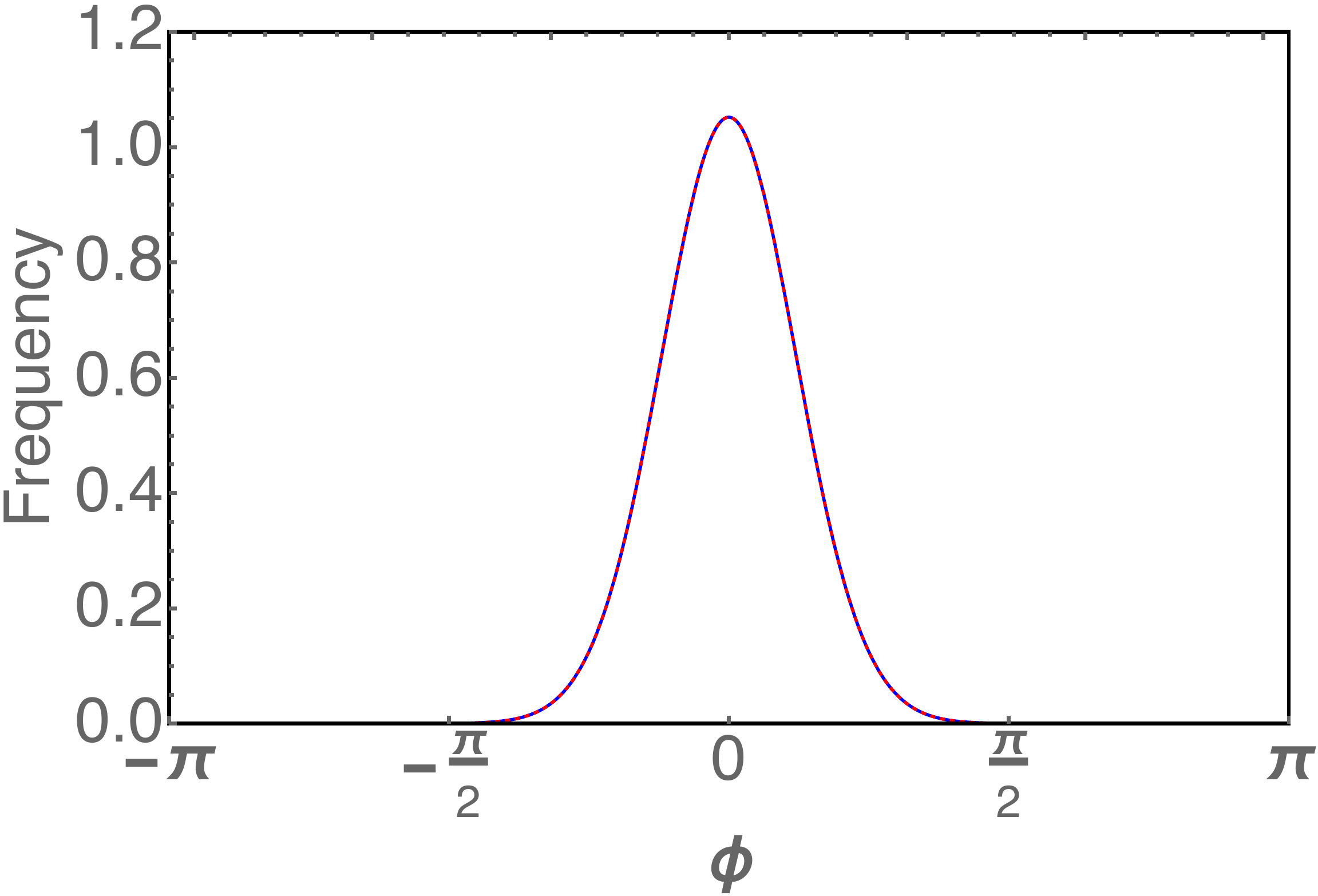}\\
\vspace{2mm}
\includegraphics[width=3.2in]{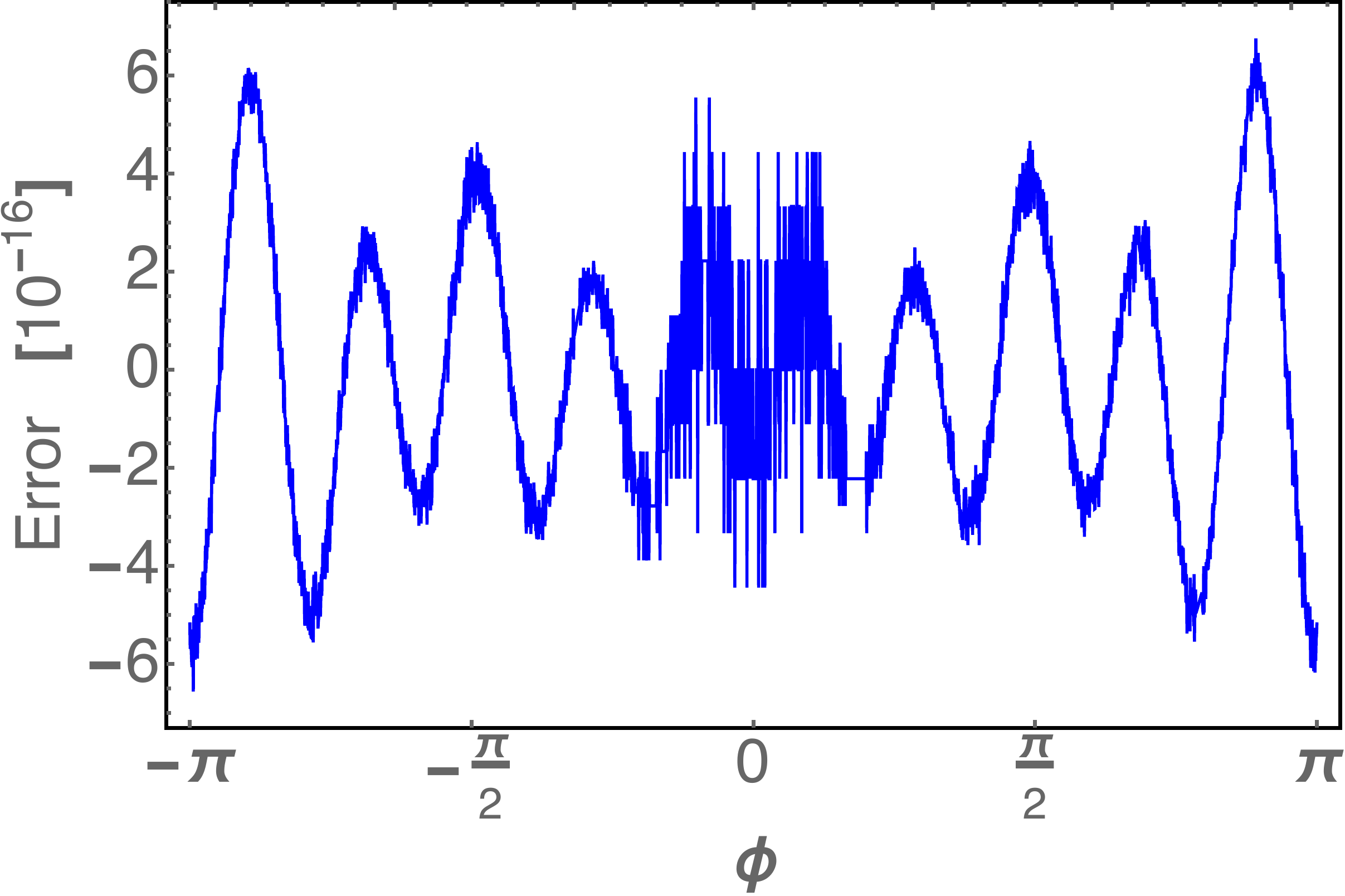}
\caption{Top: the convolution of three von Mises distributions, truncated at tenth order (dotted, red), the same with $\tilde{I}_4(z)$, $\tilde{I}_5(z)$, $\tilde{I}_6(z)$, $\tilde{I}_7(z)$, $\tilde{I}_8(z)$, $\tilde{I}_9(z)$, and $\tilde{I}_{10}(z)$ approximating the higher order Bessel functions in the computation of equation (\ref{eq:realspace}) (solid, blue). Bottom: the error defined as the difference between the distributions computed using the true values of $\tilde{I}_\nu (z)$ and the approximations given by $\tilde{I}_\nu (z)$. The von Mises parameters used are $(\kappa_1,\kappa_2,\kappa_3)=(50,50,10)$ and $(\mu_1,\mu_2,\mu_3)=(\pi,0,0)$. As seen in the bottom plot, the error in using this approximation is insignificant.}
\label{fig:2ndapprox}
\end{figure}

This construction can be repeated for higher order Bessel functions. In Table 1, we provide the coefficients for Bessel functions up to $I_{10}$. Figure \ref{fig:2ndapprox} shows this scheme applied to the closure phase distribution at $(\kappa_1,\kappa_2,\kappa_3)=(50,50,10)$. As seen in the bottom plot of Figure \ref{fig:2ndapprox}, the error in using this approximation is insignificant.

\end{document}